# Performance of the NuTeV Fe-Scintillator Sampling Calorimeter and Implications for Thin Calorimeters

S. Avvakumov[g], T. Adams[d], A. Alton[d], L. de Barbaro[e], P. de Barbaro[g], D. Berlin[g,e], R. H. Bernstein[c], A. Bodek[g], T. Bolton[d], J. Brau[f], D. Buchholz[e], H. Budd[g], L. Bugel[c], J. Conrad[b], R. B. Drucker[f*], R. Frey[f], J. Formaggio[b], J. Goldman[d], M. Goncharov[d], D. A. Harris[g], R. A. Johnson[a], S. Koutsoliotas[b], J. H. Kim[b], G. K. Krishnaswami[g], M. J. Lamm[c], W. Marsh[c], D. Mason[f], C. McNulty[b], K. S. McFarland[c,g], D. Naples[d], P. Nienaber[c], A. Romosan[b], W. K. Sakumoto[g], H. Schellman[e], M. H. Shaevitz[b], P. Spentzouris[b], E. G. Stern[b], B. Tamminga[b], M. Vakili[a], A. Vaitaitis[b], V. Wu[a], U. K. Yang[g], J. Yu[c] and G. P. Zeller[e]

[a] University of Cincinnati, Cincinnati, OH 45221
[b] Columbia University, New York, NY 10027
[c] Fermi National Accelerator Laboratory, Batavia, IL 60510
[d] Kansas State University, Manhattan, KS 66506
[e] Northwestern University, Evanston, IL 60208
[f] University of Oregon, Eugene, OR 97403
[g] University of Rochester, Rochester, NY 14627

NuTeV is a neutrino-nucleon deep inelastic scattering experiment at Fermilab. The NuTeV detector is a traditional heavy target neutrino detector which consists of an iron/liquid scintillator sampling calorimeter followed by a muon spectrometer. The calorimeter response to hadrons, muons and electrons has been measured in an *in situ* calibration beam over the energy range from 4.5 to 190 GeV. The small non-linearity of the response to hadrons is compared to the expectation from the measured ratio of responses between electrons and hadrons combined with the energy dependence of the fractional electromagnetic energy deposition in the form of neutral pions in hadronic showers $f_{\pi^0}(E_\pi)$. The predictions use $f_{\pi^0}(E_\pi)$ from the Monte Carlo simulations by GHEISHA, GFLUKA and GCALOR and also from the parameterizations of Wigmans and Groom. In addition, a study based on the NuTeV hadron calibration data of the effectiveness of a thin calorimeter is presented. The results of this study have important consequences for the energy resolution of calorimeters used in other applications; for example, measuring the cosmic ray flux in space or with balloon-based experiments.

## 1. Introduction

NuTeV is a neutrino-nucleon deep inelastic scattering experiment that ran at Fermilab during the 1996–1997 fixed target run. Some of the goals of NuTeV include a measurement of the weak mixing angle, a study of neutrino-nucleon structure functions and a search for neutrino oscillations. A neutrino interaction in the detector consists most often of a hadronic shower and sometimes an outgoing muon, whose direction and momentum can be measured.

The energy calibration of the detector is extremely important for both the weak mixing angle and structure function analyses. It represents one of the largest systematic uncertainties in those measurements. For this reason, NuTeV ran with an *in situ* calibration beam line, which was used to take data during the same accelerator cycle as the neutrino beam.

We start with brief description of the detector, followed by a description of the technique used for calorimeter calibration and the results. At the end of this paper we report a study of the effectiveness of thin calorimeters, based on the NuTeV calibration data and Monte Carlo studies.

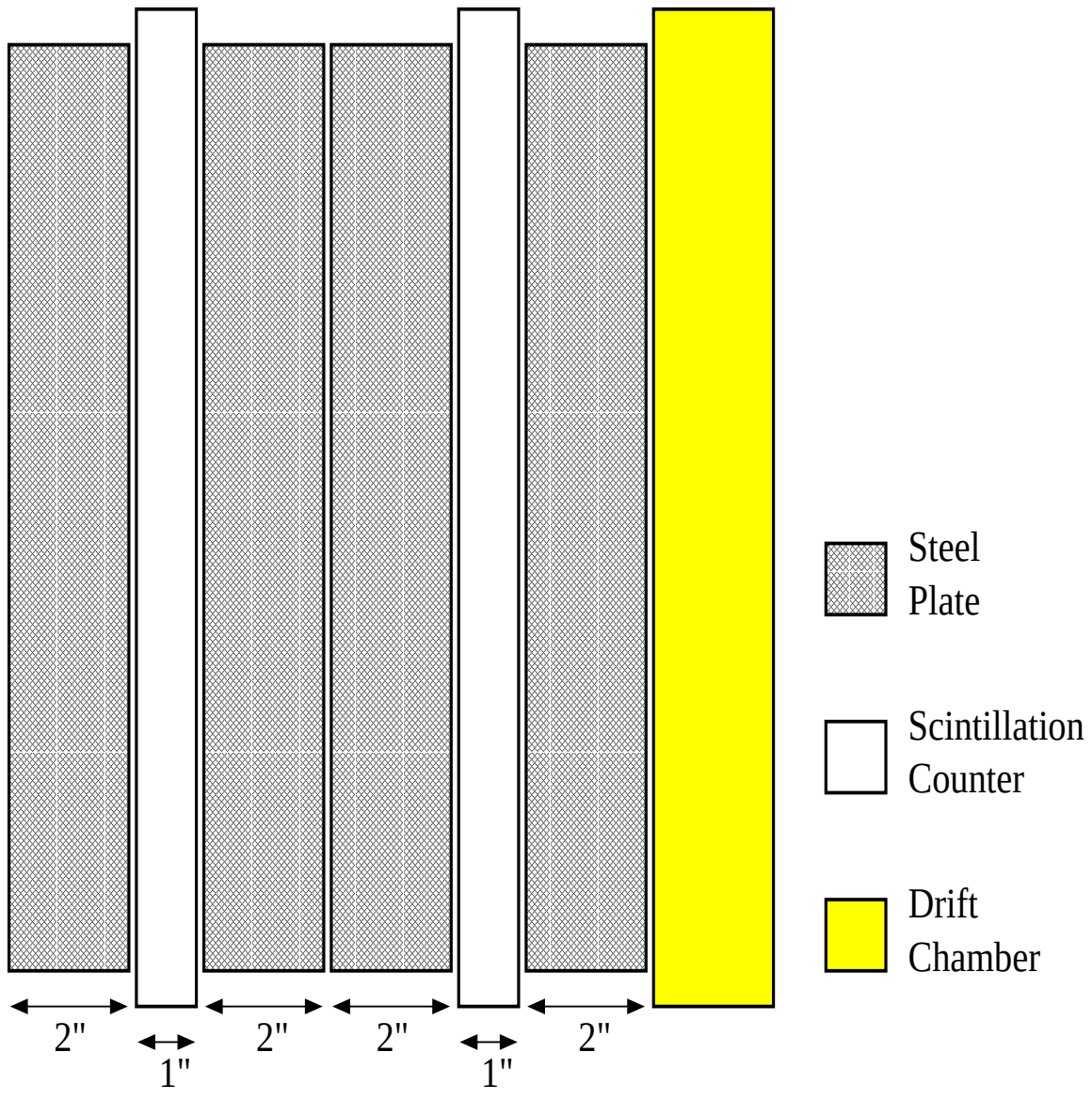

Figure 1. Geometry of one unit of the calorimeter. This unit is repeated 42 times.

| Component | Length | | |
|---|---|---|---|
| | (cm) | $\lambda_{rad}$ | $\lambda_I$ |
| 4 Steel Plates | 20.7 | 11.75 | 1.24 |
| 2 Scint. Counters | 13.0 | 0.51 | 0.16 |
| 1 Drift Chamber | 3.7 | 0.17 | 0.03 |
| Dead Space (air) | 6.0 | $2{\cdot}10^{-4}$ | $8{\cdot}10^{-5}$ |
| Total | 43.4 | 12.43 | 1.43 |

Table 1
Composition in units of cm, interaction and radiation lengths of one unit of the NuTeV calorimeter.

## 2. Calibration with muons and hadrons

The NuTeV calorimeter consists of 168 plates of steel measuring $10' \times 10' \times 2''$, interspersed with 84 scintillation counters of dimension $10' \times 10' \times 1''$ and with 42 drift chambers. There are two plates of steel between every two consecutive scintillaton counters, and one drift chamber between every other set of counters. The geometry of one unit of the target is shown in figure 1. This unit is repeated 42 times to make up the entire calorimeter. Table 1 summarizes the materials and their radiation and interaction lengths for one unit of a calorimeter longitudinal layer.

The scintillation counters are lucite boxes filled with Bicron 517L scintillator oil. The counters have $1/8''$ thick and $1''$ wide vertical ribs (which are typically spaced every 2 inches) for structural support. Each counter is surrounded by eight $5'$ longwave-shifter bars, doped with green BBQ fluor. Light is detected at the corners with four $2''$ photomultiplier tubes (Hamamatsu R2154 [1]). This particular geometry and readout scheme yields a minimum ionizing particle signal of about 30 photoelectrons.

The response of the calorimeter to muons is measured using muons from upstream neutrino interactions. Figure 2 shows a typical energy deposition for muons traversing a particular counter. There are on average 30 photoelectrons per minimum ionizing particle per counter. The truncated mean of this distribution [2] (for 77 GeV muons) is taken as 1 $Mip$ — minimum ionizing particle energy deposition. The truncated mean is determined by calculating the mean of the distribution, using all events; then taking the mean again, but only including the events between 0.2 and 2 times the previous mean. This procedure is repeated until the difference in the truncated mean between two consecutive iterations is less than 0.1%. The sample of muons used for this measurement is corrected on an event-by-event basis for the muon's momentum (if different from 77 GeV), and for the angle with respect to the direction perpendicular to the counter.

The hadron calibration beam is used to set the absolute energy scale of the experiment (relative to the muon response). The calibration beam transports particles (hadrons, muons or electrons, depending on the mode) with energies from 4.5 GeV to 190 GeV via a separate beamline routed around the neutrino beam. Calibration data are taken during a period separated by 1.4 seconds in time from the neutrino beam. The calibration beamline is instrumented with a low mass spectrometer which can reconstruct particle momenta with a 0.3% uncertainty.

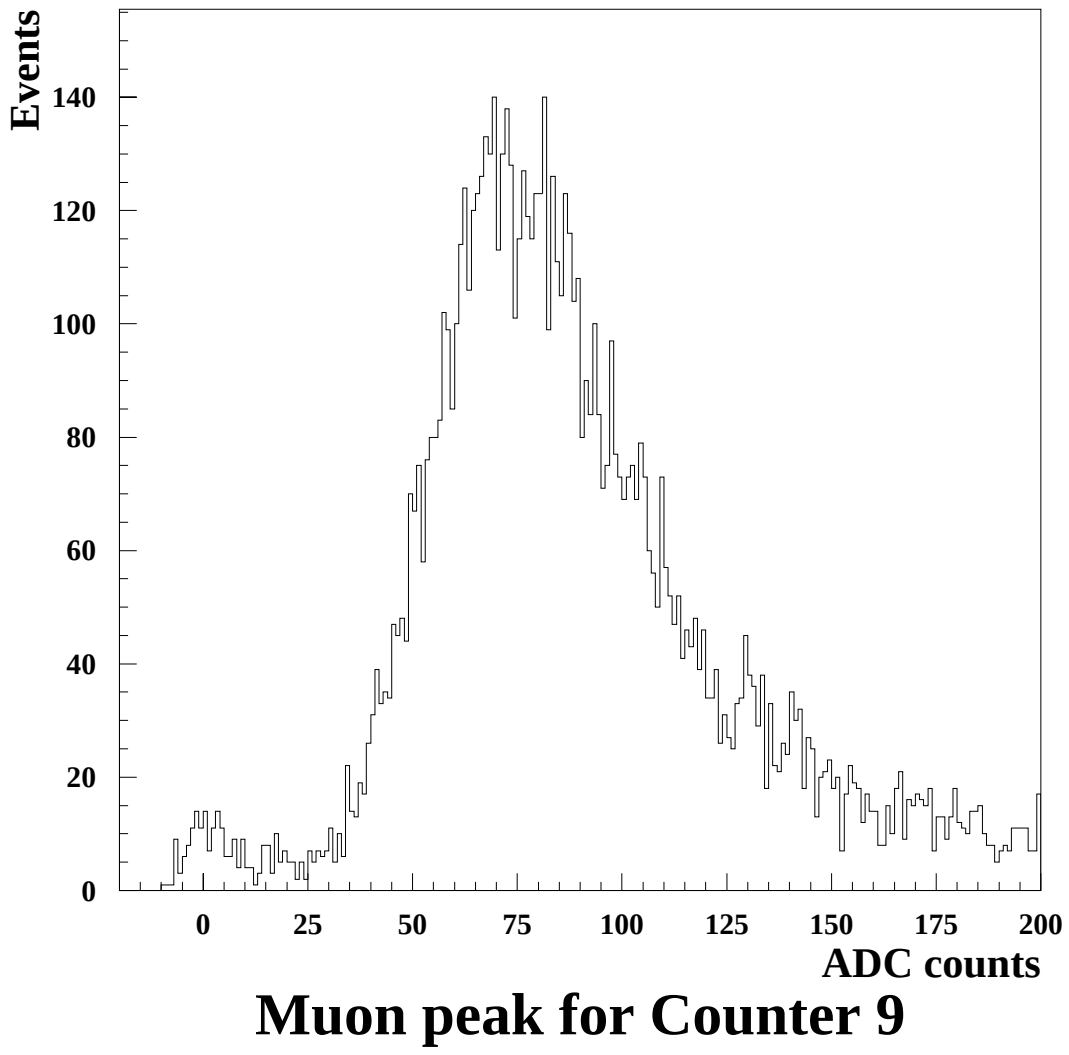

Figure 2. Typical energy deposition of muons traversing one scintillation counter.

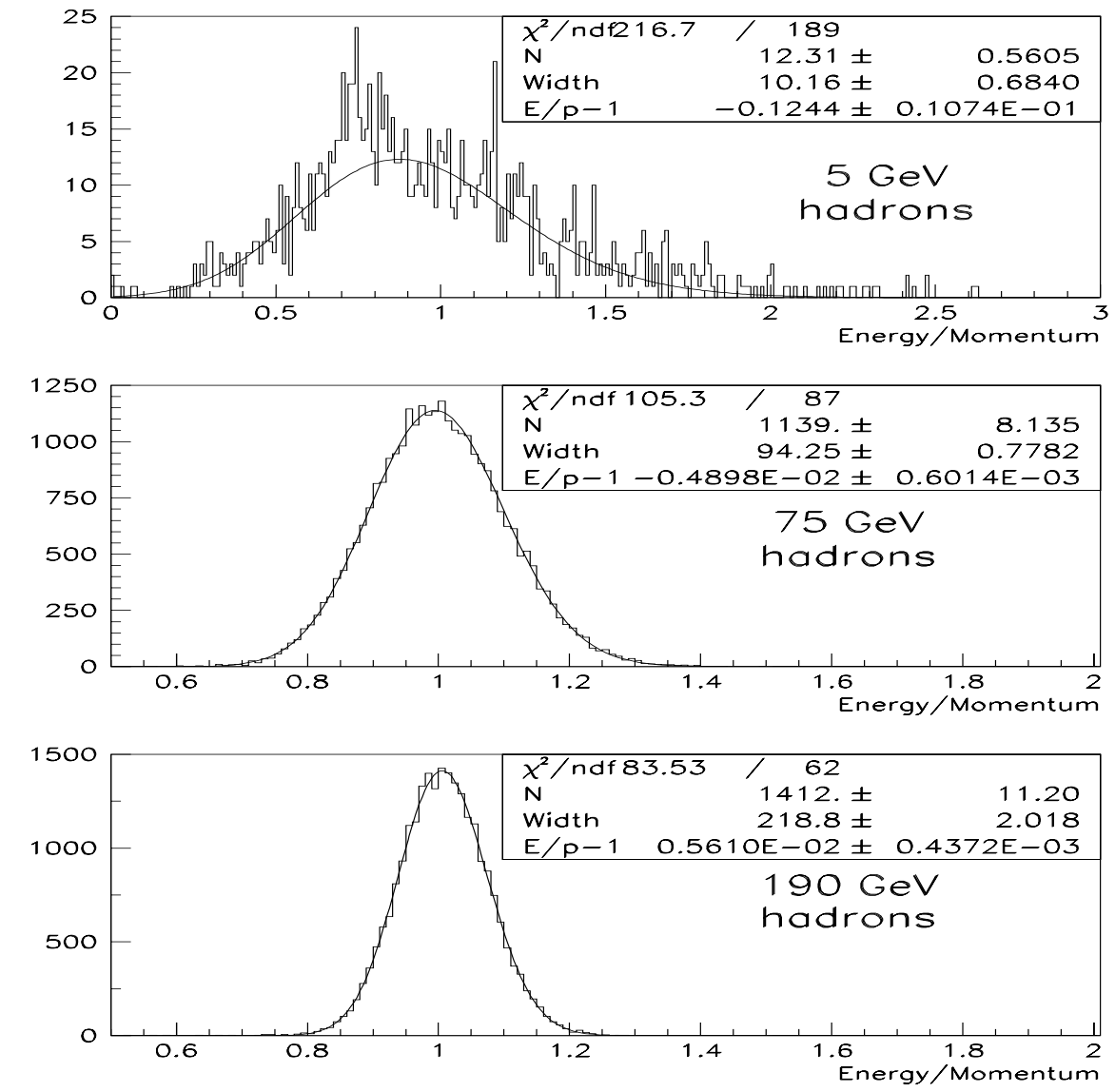

Figure 3. Poisson fits to the calorimeter energy/momentum distributions for three different energies: 4.5 GeV, 75 GeV, and 190 GeV.

In setting the hadron calibration, the total shower energy is calculated as a sum of energies deposited in the 20 most upstream counters in units of minimum ionizing particles, multiplied by a hadron calibration constant $C_\pi$:

$$E_{Shower} = \sum_{i=1}^{20} PH_i \times C_\pi \qquad (1)$$

For the NuTeV calorimeter calibration we take the initial value of $C_\pi = 0.211\ GeV/Mip$, as measured by NuTeV's predecessor experiment CCFR [3] during 1984 and 1987 Fermilab fixed target runs.

To set the absolute energy scale we fit the distributions of the measured hadron energy (E) divided by the momentum of the calibration beam particles (p) with a Poisson-like function [4]. Figure 3 shows the fit to this function for three different calibration beam modes of 4.5 GeV, 75 GeV and 190 GeV.

The energy dependence of the mean E/p distributions is shown in Fig. 4. Note that the non-linearity of the calorimeter between 4.5 GeV and 190 GeV is 4.5%. Such a small non-linearity is characteristic of non-compensating calorimeters which have only a slightly different response to hadronic and electromagnetic showers. If $C_h$ is the calorimeter calibration constant for 'pure' hadronic showers and $C_e$ for electromagnetic showers, then the calibration constant $C_\pi$ for the 'real' hadronic showers can be expressed as:

$$C_\pi(E_\pi) = f_{\pi^0}(E_\pi) \times C_e + (1 - f_{\pi^0}(E_\pi)) \times C_h \qquad (2)$$

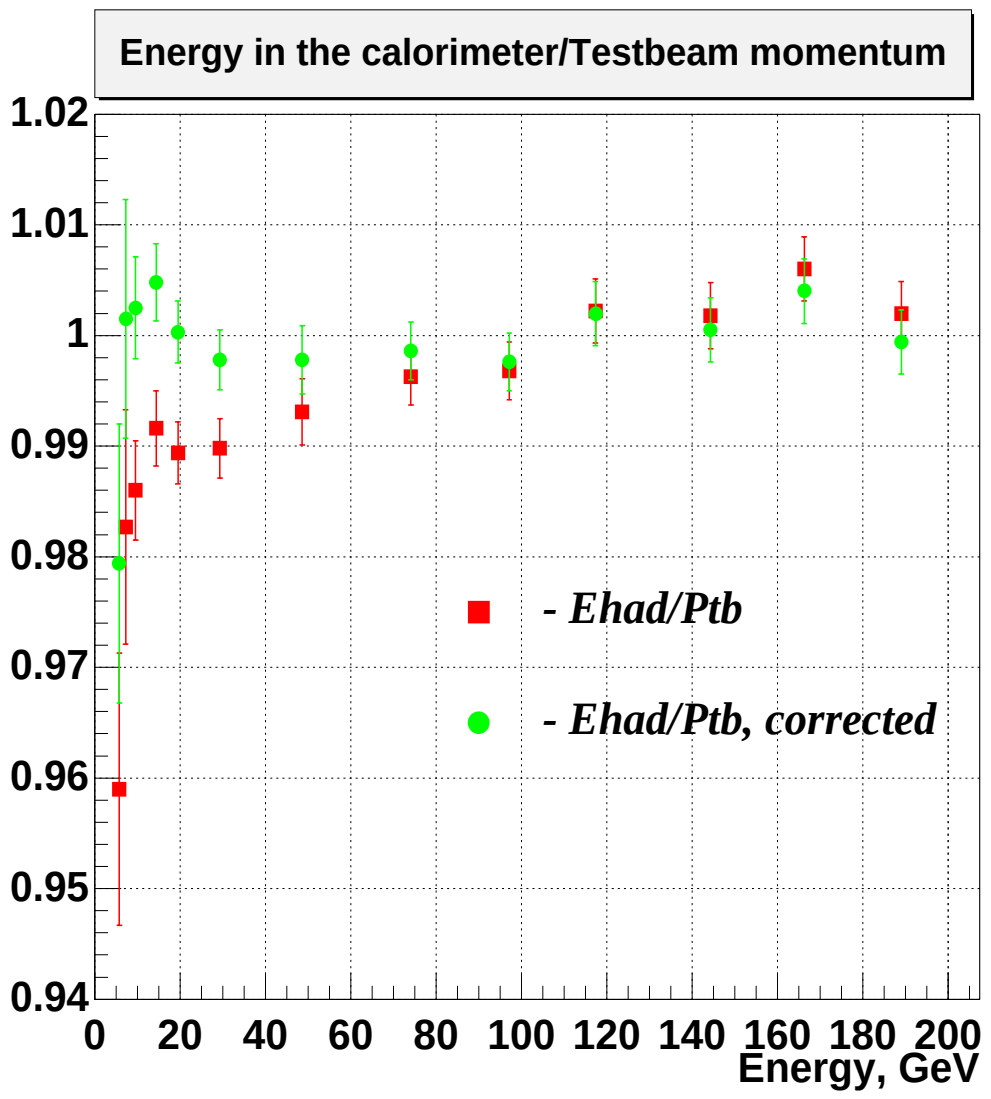

Figure 4. NuTeV calorimeter response (squares), response 'corrected' for the $C_e \neq C_h$-induced non-linearity (circles). The $f_{\pi^0}(E_\pi)$ used is from Groom's parameterization.

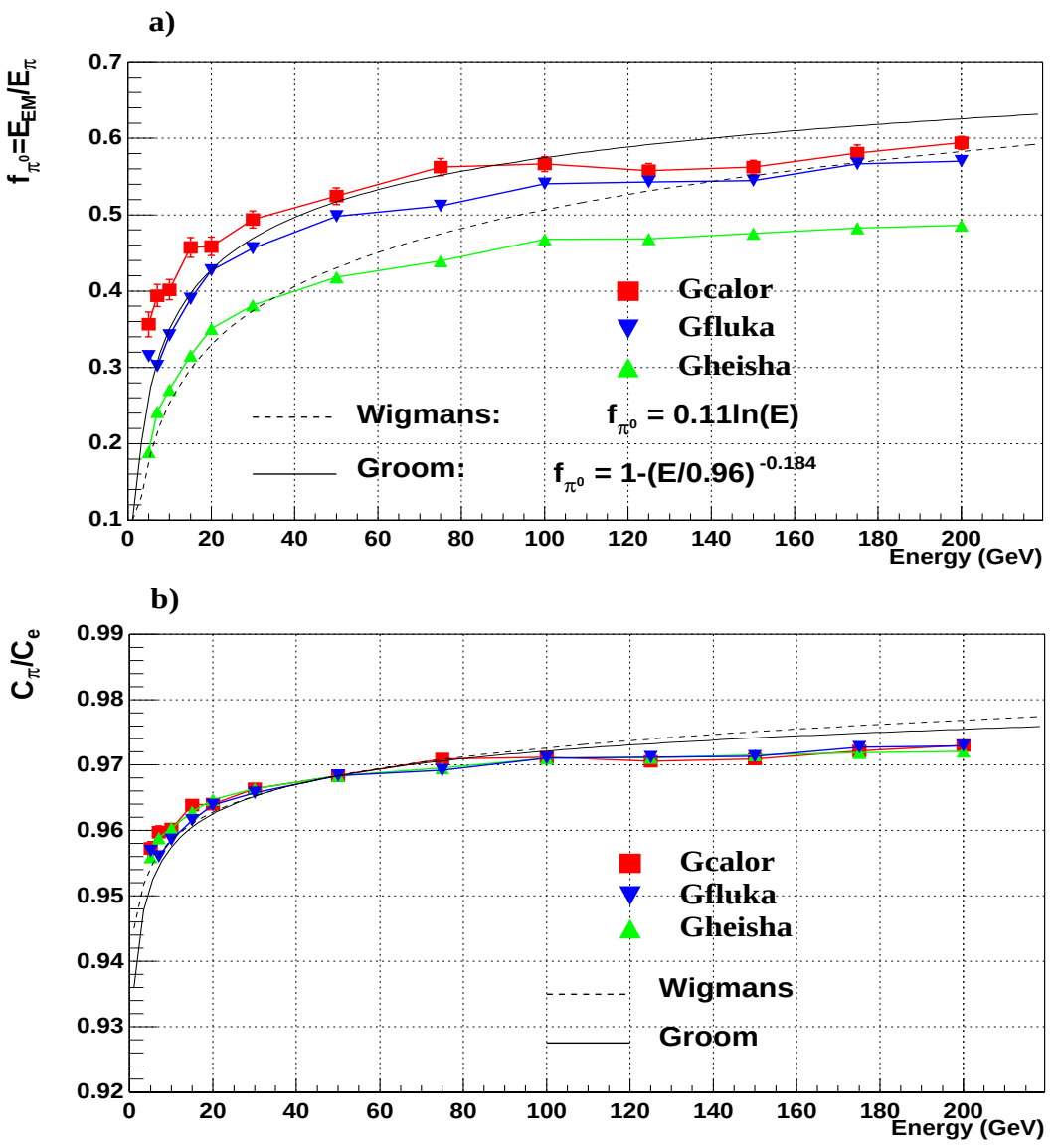

Figure 5. a) Fractional energy of the electromagnetic component in a hadronic shower as a function of hadron energy for three different hadron shower generators: GHEISHA, GFLUKA, and GCALOR, together with the Wigmans and Groom parameterizations; b) Predicted non-linearity, all three generators and both parameterizations were required to have the same value of $C_\pi$ at 50 GeV.

where $f_{\pi^0}(E_\pi)$ is the fractional energy of the initial hadron deposited through the electromagnetic process $\pi^0 \rightarrow 2\gamma$ decays. In order to predict the non-linearity of the NuTeV calorimeter we perform a GEANT [5]-based Monte Carlo simulation of the detector to determine $f_{\pi^0}(E_\pi)$ and compare the non-linearity, predicted by the equation 2 with the measured non-linearity shown in figure 4.

Monte Carlo studies have been carried out using 3 different generators commonly used within GEANT to simulate hadronic showers — GHEISHA [6] , GFLUKA [7] and GCALOR [8]. Note that only GHEISHA is native to the GEANT program. GFLUKA and GCALOR are somewhat imperfect [9] implementations of the original FLUKA and CALOR codes into the GEANT framework (and therefore can yield somewhat different results than the original FLUKA and CALOR programs). The estimated $f_{\pi^0}(E_\pi)$ from these simulations is presented in figure 5a) together with two phenomenological parameterizations of $f_{\pi^0}(E_\pi)$ by Wigmans [10] and Groom [11]. Note the Groom's parameterization of $f_{\pi^0}(E_\pi)$ is the most recent and is supposed to be a parameterization of results from the original CALOR Monte Carlo.

By requiring the three generators shown in figure 5a to have the same $C_\pi$ at 50 GeV one can construct the expected non-linearity of the detector response to hadrons as a function of en-ergy as shown in figure 5b. Although the fractional electromagnetic component as a function of energy is different for the three generators, the predicted non-linearity for hadrons is similar. By fitting the measured hadron energy response shown in figure 4 assuming a non-linearity predicted by Groom's parameterization of $f_{\pi^0}(E_\pi)$ in equation 2 we find the value of $C_e/C_h$ to be $1.053 \pm 0.01$[1], which implies a 'real' hadron to electron response ratio of $C_e/C_\pi = 1.03$ at 75 GeV. This is in agreement with the value of $1.047 \pm 0.02$ measured in the calibration beam with 75 GeV hadrons and electrons.

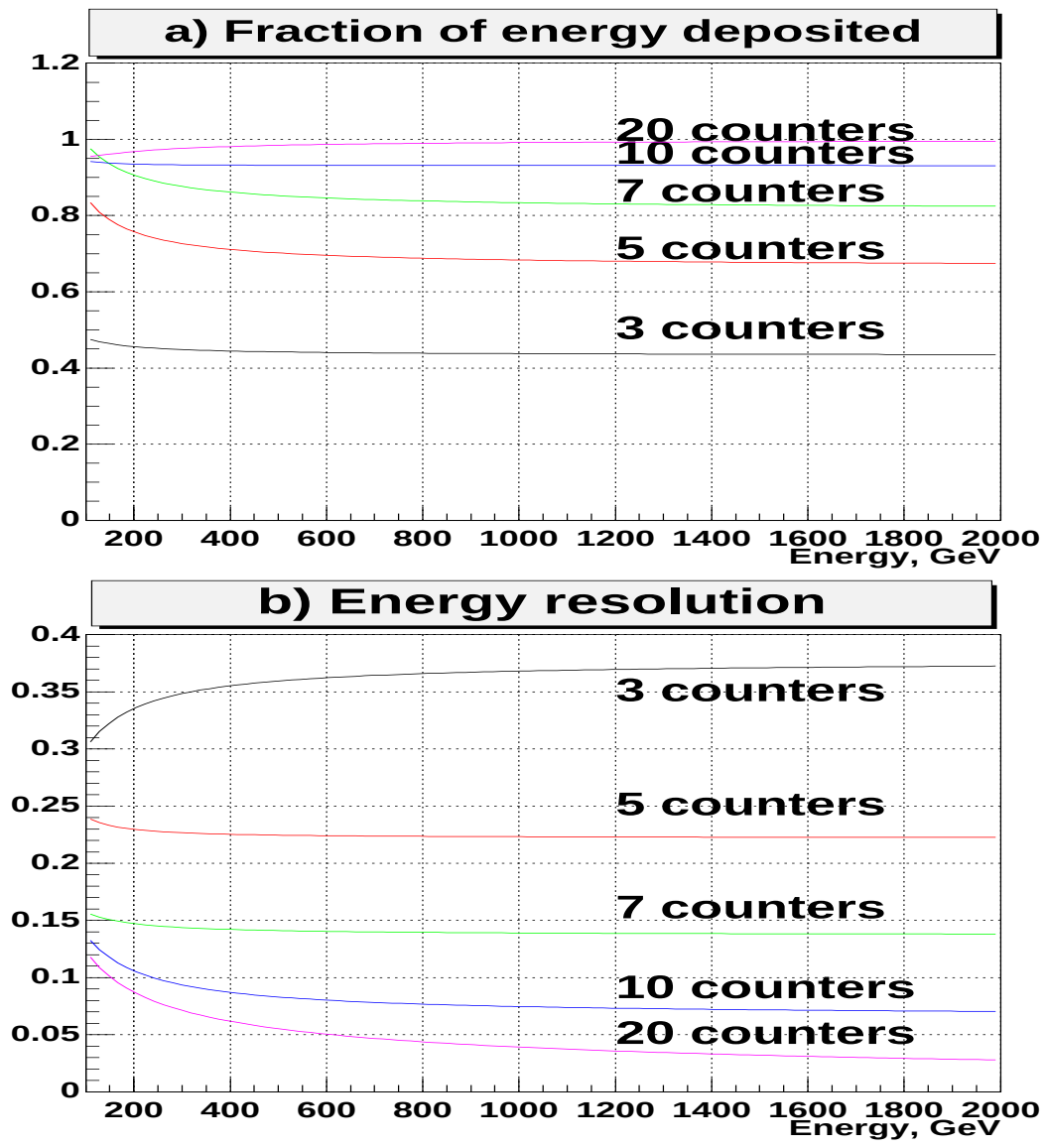

Figure 6. a) Fraction of the shower energy deposited in hadron calorimeters of various lengths; b) Energy resolution.

The hadron energy response of the NuTeV calorimeter (divided by the predicted non-linearity using Groom's parameterization for $f_{\pi^0}(E_\pi)$) is shown in figure 4. It appears that this model describes the non-linearity of the NuTeV calorimeter response to hadrons very well. A similar conclusion has been reported by the CDF Collaboration [12] for their plug-upgrade hadron calorimeter, which has a much larger non-linearity (i.e., it has a larger difference in its response for hadrons and electrons). Additional details on the calibration and response of the NuTeV calorimeter are reported in a longer communication [4].

## 3. Thin calorimetry

For possible future space-based HEP experiments (e.g., a measurement of cosmic ray hadrons) one needs to minimize the size and mass of the detector. Therefore, a natural question arises — how thin can a hadron calorimeter be made and still remain effective? A study, based on the NuTeV calibration beam data and the detector Monte Carlo, has been done to determine the energy resolution of thin calorimeters ranging from 2 to 15 nuclear interaction lengths. A very thin calorimeter of $\approx 2$ nuclear interaction lengths is of the most interest, since mass is so expensive to place in space.

We investigate thin calorimeters [13] by using only 3, 5, 7 or 10 of the 84 scintillation counters (3 counters correspond to approximately 1.8 nuclear interaction lengths). Note that there is only $2''$ of steel immediately upstream of the first counter, and $4''$ of steel immediately upstream of each of the other counters. All events are required to have an energy deposition greater than 10 GeV in the first counter to guarantee that the shower starts in the very beginning of the detector (i.e. in the first $2''$ of steel). First, we verify the validity of the Monte Carlo simulation of the detector using the calibration beam data for energies up to 190 GeV. We then use the Monte Carlo simulation to predict the energy resolution of various length calorimeters at higher energies. Using the hadron shower generator GHEISHA we are able to reproduce the measured detector resolution to within 15% (of its value) for thin calorimeters [13].

[1] Groom's parameterization predicts slightly higher non-linearity compared to the Monte Carlo generators as seen in figure 5b and we think that this fit somewhat underestimates value of $C_e/C_h$.

The results of the study of thin calorimeters are presented in figure 6. We find that with a 1.8 nuclear interaction length calorimeter the energy is measured with a constant resolution of approximately 35% to 37% with the efficiency of the imposed '10 GeV in the first counter' cut increasing from 15% at 100 GeV to 30% at 2 TeV. The fraction of the interaction energy which is contained in 3 counters is approximately 43%. The energy in the first 1.8 interaction length is almost entirely from the electromagnetic component of the first interaction of the high energy hadron. Additional details are reported in a longer report [13].

## 4. Summary

In conclusion:

1. The NuTeV calorimeter has been calibrated using neutrino-induced muons, which can be used to account for time and position dependence of the calorimeter response.

2. The absolute energy scale has been established using the NuTeV hadron calibration beam with a precision of 0.3% [4].

3. A 4.5% non-linearity in the hadron response between 4.5 and 190 GeV has been measured, and is found to be consistent with the prediction using Groom's parameterization of $f_{\pi^0}$, with $C_e/C_h = 1.053 \pm 0.01$.

4. The hadron energy resolution [4] of the NuTeV calorimeter is:

$$\frac{\sigma}{E} = (.020 \pm .002) + \frac{(.885 \pm .004)}{\sqrt{E}}$$

5. The response and resolution of thin calorimeters with 1.8–10 nuclear interaction lengths have been measured for energies up to 200 GeV, and simulated up to 2 TeV. The energy resolution of a 2 interaction lengths calorimeter is predicted to be constant at a level of 35% at high energy [13].

**Acknowledgements**

We would like to express our gratitude to the US Department of Energy and the National Science Foundation for their support. We thank the numerous people at Fermilab and the collaborating institutions who helped in all the facets of this project.